\documentclass[aps,10pt,a4paper,showpacs,jcp,twocolumn]{revtex4-1}
\usepackage{amssymb}
\usepackage{mathrsfs}
\usepackage{stmaryrd}
\usepackage{subfigure}

\usepackage{graphicx}
\usepackage{color}

\usepackage{fancyhdr}
\newcommand{\mb}{\mathbf}
\newcommand{\mc}{\mathcal}

\newcommand{\mr}{\mathrm}
\newcommand{\tb}{\textbf}

\begin{document}

\title{Hybrid Particle-Continuum Simulations Coupling Brownian
Dynamics and Local Dynamic Density Functional Theory}

\author{Shuanhu Qi}
\affiliation{Institut f\"{u}r Physik, Johannes Gutenberg-Universit\"{a}t Mainz, Staudingerweg 7, D-55099 Mainz, Germany}

\author{Friederike Schmid}
\affiliation{Institut f\"{u}r Physik, Johannes Gutenberg-Universit\"{a}t Mainz, Staudingerweg 7, D-55099 Mainz, Germany}

\bigskip

\begin{abstract}

We present a multiscale hybrid particle-field scheme for the simulation of
relaxation and diffusion behavior of soft condensed matter systems. It combines
particle-based Brownian dynamics and field-based local dynamics in an adaptive
sense such that particles can switch their level of resolution on the fly. The
switching of resolution is controlled by a tuning function which can be chosen
at will according to the geometry of the system. As an application, the hybrid
scheme is used to study the kinetics of interfacial broadening of a polymer
blend, and is validated by comparing the results to the predictions from pure
Brownian dynamics and pure local dynamics calculations.

\end{abstract}

\maketitle

\newpage

\section{Introduction}

Dynamical processes in soft materials are typically characterized by multiple
time and length scales, an intrinsic feature that brings significant challenge
for the design of efficient simulation schemes
\cite{multiscale_bridge,multscale_rev}. Although the computational power still
continuously increases each year, studying macroscopic properties of polymer
systems by atomistic simulations has remained infeasible. Therefore, to study
processes on larger time and length scales, coarse-grained (CG) models have
been developed as well as multiscale modelling schemes where different length
scales are coupled concurrently.  Among these simulation techniques, hybrid
adaptive resolution schemes are particularly useful, since in such schemes most
parts of the system are represented by a lower resolution model, to save time,
while the detailed description holds only where it is required, to keep
accuracy.  Depending on the concrete representation schemes, the hybrid models
may be categorized into hybrid particle-particle methods (HPP)\cite{AM_CG1,
AM_CG_H, AM_CG_H1, AdResS1, QM_CG1, QM_CG2,Adaptive_reservoir, H_AdResS}, where
both the fine-grained and coarse-grained description are constructed using
particle-based models, and hybrid particle-continuum (or particle-field) (HPF)
methods, where the fine-grained molecules are resolved by particles while the
coarse-grained molecules are represented by some continuum quantities which are
controlled by some continuous dynamic equations
\cite{HybridMD1,HybridMD2,triple_scale, MD_FH1,MD_FH2,HPF1,HPF2}.  Well-known
examples of HPP schemes are the hybrid QM/MM methods \cite{QM_MM} or the hybrid
atomistic-CG approaches \cite{AM_CG1, AM_CG_H, AM_CG_H1, AdResS1}. HPF schemes
include hybrid particle-density functional schemes \cite{HPF1,HPF2} and hybrid
schemes linking molecular dynamics and fluctuating hydrodynamics
(MD/FH)\cite{HybridMD1, HybridMD2, MD_FH1, MC_NS}.  However, so far, the HPF
schemes have either been restricted to static properties - such as, e.g., the
Monte Carlo scheme proposed by us in previous work \cite{HPF1} - or to fluids
of small molecules - such as, e.g., the MD/FH schemes where molecular fluids
are connected to Navier-Stokes fluids via buffer zones\cite{HybridMD1, MD_FH1,
MD_FH2, MD_SDPD1, MD_SDPD2, MD_MCD}.

In the present paper, the HPF will refer to adaptive resolution models where
materials are represented at different levels -- particle or field -- in
different spatially separated domains, as explained above.  However, we should
note that in the literature, the term ``hybrid particle-field modeling'' is
also used for a variety of other, intrinsically different models:

Originally, the idea of hybrid particle-field modeling was exploited for the
study of two-component systems with one component described by a particle-based
model while the other uses a field model. For example, in the late 90s, Miao et
al \cite{brush_shear} and Saphiannikova et al \cite{brush_NS} proposed a mixed
scheme to study the conformational properties of polymer brushes under shear.
They utilized the continuous Brinkman equation (a variant of the Navier-Stokes
equation) to describe the motion of the solvent, and Brownian dynamics (BD) to
simulate polymer brushes. Similar ideas were later adopted by Sides et al
\cite{HPF_Sides}, with the difference that they focussed on a colloid-polymer
composite, and treated polymers as continuous fields and colloids as Brownian
particles. The main feature of these hybrid approaches is that different
resolution schemes are chosen for specified components at the very beginning,
and these representations do not change during the simulation.

Ganesan and coworkers \cite{BD_DMFT, BD_DSCF} and Milano and Kawakatsu
\cite{Milano1, Milano2, Milano3} introduced another type of single-resolution
``hybrid particle-field modeling'', where the fields are essentially used to
define the interactions between particles. The whole simulation is controlled
by a particle-based BD or MD algorithm, while the fields, e.g., densities,
appear as intermediate variables, i.e., they are extracted from all the
particle positions and then used for the calculation of interaction forces. The
introduction of these fields can be viewed as a way to efficiently deal with
nonbonded molecular interactions. A similar approach was first proposed by
Laradji et al \cite{Laradji,off_lattice}, and also characterizes the ``single
chain in mean field theory" \cite{SCMF1,SCMF2}.

In contrast, the MD/FD approach first proposed by de Fabritiis et al
\cite{HybridMD1, HybridMD2} is a ''true'' HPF model with adaptive resolution,
which which combines particle-based MD simulations and continuum-based
fluctuating hydrodynamic models dynamically. In this model the stochastic
equations are integrated by numerical methods of computational fluid dynamics.
Mass and momentum conservations are achieved through flux balance restriction.
Later, this HPF scheme was combined with an HPP (atomistic/CG) model leading to
a triple-scale approach \cite{triple_scale}. Similar MD/FD schemes using
other Navier-Stokes solvers have been proposed, e.g., methods based on smoothed
dissipative particle dynamics \cite{MD_SDPD1, MD_SDPD2} or on multiparticle
collision dynamics \cite{MD_MCD}.

Recently, we have developed a different HPF method which is specifically
designed for large, complex molecules, i.e., polymers. It couples a particle
region to a field region where the thermodynamics is described by polymer
density functional theory \cite{HPF1, HPF2}. The simulation algorithm uses 
a combination of Monte Carlo (MC) moves and field relaxation steps. We have
shown how to construct such hybrid models systematically by field-theoretic
methods.  The derivation is based on the fact that the particle
representation and field representation originate from the same partition
function obtained from the same Edwards Hamiltonian \cite{Edwards1, Edwards2,
Edwards3, Edwards4}. Thus in principle, the particle description, the field
description, and the derived hybrid description are almost equivalent and
should give the same macroscopic thermodynamic properties. This equivalence has
been verified within statistical errors through the study of polymer
conformation properties in a solution \cite{HPF2} and effective interactions
between two polymer grafted colloids immersed in a diblock copolymer melt
\cite{HPF1}. However, so far the method cannot be used to study dynamical
processes.

In this paper, we extend our previous static model and move one step forward to
construct a dynamic hybrid HPF scheme for the simulation of systems with
diffusive dynamics. The scheme ensures local mass conservation, but it neglects
inertia effects and thus does not account for momentum conservation. Therefore,
it is suited for studying relaxational dynamics and diffusion behavior in
systems where hydrodynamic interactions are not important. More specifically,
particles move according to Brownian dynamics (BD), and fields are propagated
according to the equations of local dynamic density functional theory (LD)
\cite{LD1, LD2, LD3}. Polymers can switch between different levels of
resolution on the fly.  These ''resolution switches'' are controlled by a
chemical potential like function, i.e., a tuning function, rather than by a
force or potential interpolation as in other adaptive resolution models. The
switch process is constructed such that local mass conservation is guaranteed.
We illustrate the method at the example of a study of interface broadening in
an A/B polymer blend.

The remainder of the paper is organized as follows: In Sec \ref{sec:method}, we
describe our methodology at the example of a polymer solution with implicit
solvent.  The three components of our hybrid scheme, the BD method
(\ref{subsec:bd}), the LD method (\ref{subsec:ld}) and identity switches
(\ref{subsec:switch}) are discussed separately. In Sec \ref{sec:results}, we
present simulation results. We first test the constructed hybrid scheme using
the polymer solution and then utilize it to study the interdiffusion process of
a polymer blend. We close with a brief summary in Sec \ref{sec:summary}.

\section{Hybrid Particle-Field Scheme: Methodology}
\label{sec:method}

In this section we describe in detail the construction of the HPF model. For
simplicity, we focus on a solution of homopolymer Gaussian chains. The
extension to more complicated polymers or copolymers and to mixtures etc. 
is straightforward. 

We consider a system containing $n_\mr t$ Rouse polymers of length $N$
confined in a cubic box of volume $V=L_\mr x\cdot L_\mr y\cdot L_\mr z$ with
periodic boundary conditions. In the following and throughout the paper, we
give all lengths in units of the radius of gyration of a free Gaussian chain
$R_\mr g=\sqrt{Nb^2/6}$, where $b$ is the statistical Kuhn length, energies in
units of $k_BT$, where $k_B$ is the Boltzmann constant and $T$ the temperature,
and times in units of the relaxation time of a single chain, i.e., $t_0=R^2_\mr
g/D_\mr c$, where $D_\mr c$ is the mobility of the center of mass of a single
chain. The original Hamiltonian of the system is written as the sum of two
contributions $\mc H=\mc H_0+\mc H_\mr I$, where $\mc H_0$ accounts for the
chain connectivity and the bonded intrachain interactions, and $\mc H_\mr I$
represents the nonbonded interactions between chain monomers.  Written in terms
of local densities, it has the form \cite{Edwards1, Edwards2, Edwards3,
Edwards4} 
\begin{equation}
\label{eq:HI}
\mc H_\mr I= \frac{n_\mr tv_\mr{ex}}{2V}\int d\mb r \hat\phi^2,
\end{equation}
where $v_\mr{ex}$ is the excluded volume parameter, and $\hat\phi(\mb r) \equiv
\frac{1}{\rho_0}\sum_\mr{mj}\delta(\mb r-\mb R_\mr{mj})$ is the normalized
microscopic density with respect to the total mean monomer density
$\rho_0=n_\mr tN/V$.

In the HPF scheme, the chains are first partitioned into ''particle type
chains'' (p-chains) and ''field type chains'' (f-chains), depending on the
state of virtual internal degrees of freedom as described below in Sec
\ref{subsec:switch}. The f-chains are then turned into fluctuating density
fields by a field-theoretic transformation and a saddle-point
integration\cite{Hong, SCF1, HPF1, HPF2}.  This results in an effective free
energy functional of the form
\begin{equation}
\label{eq:ftotal}
	\mc F[\{\mb R_\mr{mj}\}, \phi_\mr f]
    = \mc F_{0,\mr p}[\{\mb R_\mr{mj}\}] 
        + \mc F_{0,\mr f}[\phi_\mr f] 
        + \mc F_\mr I[\{\mb R_\mr{mj}\},\phi_\mr f]
\end{equation}
for the particle-field system, which depends on the positions $\{\mb
R_\mr{mj}\}$ of p-chain monomers -- the index m refers to the molecule, and the
index j to the monomer in the chain -- and on the local densities $\phi_f$ of
f-chain monomers.  The terms $\mc F_{0,\mr p}$ and $\mc F_{0,\mr f}$ include
the contributions of the intrachain interactions and the chain entropy of
p-chains and f-chains, respectively, and $\mc F_I$ describes the bonded
interactions, which is derived from (\ref{eq:HI}) and given by 
\begin{equation}
\label{eq:FI}
	 \mc F_\mr I=
	 \frac{n_\mr t v_\mr{ex}}{2V}\int d\mb r \: 
        (\phi_\mr f + \hat\phi_\mr p)^2.
\end{equation}
in a homopolymer solution.
Here $\hat\phi_\mr p(\mb r)$ of course refers to the normalized local density
of p-chains only.

For Gaussian chains, $\mc F_{0,\mr p}$ can be written as \cite{Doi}
\begin{equation}
\label{eq:F0_p}
	\mc F_{0, \mr p}=
	\frac{N}{4}\sum_{\mr m=1}^{n_\mr p} \sum_{j=2}^\mr N
	\big(\mb R_\mr{mj}-\mb R_\mr{m,j-1}\big)^2.
\end{equation}
The sum m runs over the p-chains and $\mb R_\mr{mj}$ is the
position of the jth bead of the mth chain. 

Finally, the intrachain contribution of f-chains to the free energy,
$\mc F_0^\mr f$, takes the general form
\begin{equation}
\label{eq:F0_f}
	\mc F_{0, \mr p}=
        -\frac{n_\mr t}{V} \int d\mb r \: \phi_\mr f \: \omega_\mr f
        - n_\mr f\ln[Q_\mr fV],
\end{equation}
where $\omega_\mr f(\mb r)$ are conjugate fields defined such that hypothetical
external potentials $W = \frac{1}{N} \omega_\mr f$ would generate exactly
the f-density configuration $\phi_\mr f(\mb r)$ in a reference system of
noninteracting f-chains, and $\mc Q_\mr f$ is the partition function of a
single chain subject to such potentials $W_\mr f$.  For Gaussian chains, it is
given by 
\begin{eqnarray}
\label{eq:Q}
    \mc Q_\mr f &=& \frac{1}{Z_0} \int \mr d \bf R_1 \cdots \mr d \bf R_N \\
     &\times& \exp\Big\{
        - \frac{N}{4} \sum_{j=2}^N (\mb R_j - \mb R_{j-1})^2
        - \frac{1}{N} \sum_{j=1}^N \omega(\mb R_j) \Big\}.
\nonumber
\end{eqnarray}
We note that in this formalism, $\phi_\mr f$ is the only free field, and
$\omega_\mr f$ must be determined from $\phi_\mr f$. An explicit relation
between $\phi_\mr f$ and $\omega_\mr f$ can be obtained through the following
procedure: We define an end-integrated single chain propagator $q_\mr f(\mb
r,s)$, which satisfies the modified diffusion equation
\cite{SCF1,SCF2}
\begin{equation}
\frac{\partial q_\mr f}{\partial s}=\nabla^2 q_\mr f-\omega_\mr fq_\mr f,
\end{equation}
where $s$ is a normalized chain contour variable ranging from 0 to 1. The
propagator of $q_\mr f$ is then calculated numerically with initial condition
$q_\mr f(\mb r,0)=1$ and periodic boundary conditions, e.g., using a
pseudo-spectral method. Knowing the propagator $q_\mr f$, one can calculate the
single chain partition function via $Q_\mr f=\frac{1}{V}\int d\mb r \: q_\mr
f(\mb r,1)$, and the density field via \cite{SCF1,SCF2}
\begin{equation}
    \phi_\mr f =
    \frac{\bar\phi_\mr f}{Q_\mr f}\int_0^1 ds \:
        q_\mr f(\mb r,s) \: q_\mr f(\mb r,1-s).
\end{equation}
The functional inversion to get $\omega_\mr f$ from $\phi_\mr f$ (density
targeting problem) is done numerically using iteration techniques. Several
iteration schemes are available; for more details, see, e.g.,
\cite{target_density, fluctuation_dynamics,stasiak}.

In the dynamic HPF scheme, the motion of the particle-resolved p-chains is
described by BD dynamics, whereas the density profiles of the coarse-grained
f-chains evolve according to a dynamic density functional, the LD scheme. The
switches between p-chains and f-chains are controlled by a predefined tuning
function. Apart from the switches, the system is propagated by numerical
integration of the BD equations of motion (p-chains) and the LD equation
(f-chains). More concretely the evolution of the system in each integration
step is composed of three independent steps. First, all the beads' positions of
p-chains are updated according to the BD equations, second, density profiles
are evolved according to the LD equations, and third, a finite number of
identity switches are carried out controlled by a Monte Carlo (MC) criterion.
An advantage of completely decoupling the updating and the switching process is
that the hybrid algorithm becomes more flexible with respect to incorporating
other dynamic models, and can be extended to other polymer models more easily.
In the following we describe the BD dynamics, the LD method, and the switching
algorithm respectively.

\subsection{Particle Propagation: Brownian Dynamics} 
\label{subsec:bd}

The motion of particle beads (in p-chains) is driven by a conservative force
derived from the Hamiltonian and a random force originating from the thermal
fluctuations, i.e.,
\begin{equation}
\frac{d\mb R_\mr{mj}}{dt}=-D_0\frac{\partial\mc F}{\partial\mb R_\mr{mj}}+\sqrt{2D_0}\mb f_\mr{mj},
\end{equation}
with $D_0=1/N$ (in units of $R^2_\mr g/t_0$). The random force $\mb f_\mr{mj}$
is Gaussian distributed with zero mean and variance $\langle
f_\mr{mjI}(t)f_\mr{nkJ}(t')\rangle=\delta_\mr{mn}\delta_\mr{jk}\delta_\mr{IJ}$
where I,J denote the Cartesian components. The derivative of Hamiltonian over
particle positions can be performed directly using the chain rule and
properties of the delta function. The result can be formally written as
\cite{slowdown}
\begin{equation}
\frac{\partial\mc H}{\partial\mb R_\mr{mj}}=\frac{N}{2}\big(\mb R_\mr{mj}-\mb R_\mr{m,j+1}-\mb R_\mr{m,j-1}\big)+\frac{1}{N}\frac{\partial\hat\omega_\mr p}{\partial\mb R_\mr{mj}},
\end{equation}
where $\hat\omega_\mr p$ can be viewed as the conjugate potential to
$\hat\phi_\mr p$, and is obtained as
\begin{equation}
\hat\omega_\mr p=v_\mr{ex}(\hat\phi_\mr p+\phi_\mr f)
\end{equation}
in our homopolymer solution.  Therefore the nonbonded force acting on one bead
is evaluated through the derivative of densities with respect to the bead's
position. 

In practice, all densities are discretized, and they are extracted by a
``particle-to-mesh" assignment function. To do so, we uniformly divide the
simulation box into $n_\mr x\cdot n_\mr y\cdot n_\mr z$ cells, and define all
densities on the vertexes (mesh points) of these cells. Several choices of the
assignment function \cite{PM} are possible. The zeroth order scheme is the so
called the nearest-grid scheme \cite{MC_brush1}, which assigns a 
bead to its nearest mesh point. Here we adopt a linear order scheme, which
assigns fractions of a bead to its eight nearest mesh points, the so-called
cloud-in-cell (CIC) scheme \cite{MC_brush2,CIC}. In the CIC scheme, the
fraction assigned to a given vortex is proportional to the volume of a
rectangular whose diagonal is the line connecting the partition position and
the mesh point on the opposite side of the cell. The assignment function
completely determines the finite-differential form of the derivative of
$\hat\omega_\mr p$. The nonbonded force acting on a bead is finally expressed
as the linear interpolation of $\hat\omega_\mr p$ defined on the bead's eight
nearest mesh points with weights calculated according to the distance between
the particle and the mesh point. As the explicit form of the assignment
function and the nonbonded force are extensively discussed elsewhere, we just
refer to Refs.\cite{slowdown,PM}.

\subsection{Field Propagation: Local Dynamic Density Functional}
\label{subsec:ld}

The density fields $\phi_\mr f$ of f-chains are propagated according to a
dynamic density functional. The starting point is the continuity equation for
the monomer densities $\rho_0 \frac{\partial \phi_\mr f}{\partial t} = - \nabla
\mb j$, where the current $\mb j$ is driven by the thermodynamic force $\nabla
\mu_\mr f$ with $\mu_\mr f :=\frac{\delta\mc F}{\delta\phi_\mr f}$
\cite{fluctuation_dynamics}.  Here we use a local dynamics scheme, where the
thermodynamic force and the flux are linearly related, $\mb j=-D_c N \rho_0 \phi_\mr f\nabla\mu_\mr f$, leading to the LD equation \cite{LD1,LD2,LD3}:
\begin{equation}
\label{eq:ld}
   \frac{\partial\phi_\mr f}{\partial t}
    = D_c \nabla\cdot\Big[
      \phi_\mr f\nabla\frac{\delta\mc F}{\delta\phi_\mr f}
     \Big]
\end{equation}
with $\delta \mc F/\delta \phi_\mr f = \frac{n_\mr t}{V}[ v_\mr{ex}
(\hat{\phi}_\mr p + \phi_\mr f) - \omega_\mr f]$ in the case of the
homopolymer soluation.

In the present work, the LD equation is integrated numerically using the
explicit Euler scheme, and the derivatives are evaluated in Fourier space using
fast Fourier transform (FFT) \cite{FFT}. To perform the numerical calculations,
we discretize all the continuum quantities and define them only on the mesh
points for convenience.  Thus the space decomposition used in the LD dynamics
is the same as that used in the BD simulation when calculating the densities.

\subsection{Chain Identity Switches}
\label{subsec:switch}

Originally, all $n_\mr t$ chains in the system are identical. For the HPF
modeling, we need to first partition these chains into p-chains and f-chains.
There are several ways to do the partitioning. The strategy we adopt here is
based on the introduction of an additional ``spin" variable $\tau_\mr{mj}$,
which is attached to each bead as an identifier \cite{HPF1,HPF2}. The spin
$\tau_\mr{mj}$ can take a value either 0 or 1. A chain is identified as an
f-chain if all the spins on this chain are set zero, otherwise it is a p-chain.
We note that this partioning method leads to the asymmetry of p-chains and
f-chains in the spin space, where p-chains have a larger spin entropy. To
control the value of $\tau_\mr{mj}$, we further introduce a tuning function
$\Delta\mu(\mb r)$, which plays the role of the spin's conjugate potential. The
tuning function couples to $\tau_\mr{mj}$, and thus determines the
distributions of p-chain and f-chains. Therefore, p-chains and f-chains are
treated as two chemically different species, and $\Delta\mu$ acts similar to a
chemical potential difference, which however varies locally. From a technical
point of view, we thus work in a semi-grand canonical ensemble.

In the present study, the tuning function is imposed externally and must be
specified beforehand. Some general considerations can help to construct an
appropriate form for $\Delta\mu$. For example, it should be chosen such that
p-chains appear only in specific regions (e.g., near boundaries), where a
detailed description is required, while in the remaining large region, chains
are represented by coarse-grained fields. Based on such considerations, an
optimal form of $\Delta\mu$ may be constructed that it is in compatible with
the geometry of the localized region as well as the whole system. 

The chain identity switches are driven by the tuning function, and to sample
the spin variable, we adopt a Monte Carlo type scheme \cite{book_MC1,book_MC2}.
It is motivated by the following consideration: We require that a spin variable
$\tau$ on a bead at position $\mb r$ assumes the value $\tau = 1$ with a
probability $w_1$ roughly given by \cite{HPF1, HPF2}
\begin{equation}
\label{eq:target}
	w_1=\frac{e^{\Delta\mu(\mb r)}}{1+e^{\Delta\mu(\mb r)}}
\end{equation}
while the probability for $\tau=0$ is $w_0=1-w_1$.  In the {\em absence} of
other dynamical processes (BD or LD moves), this probability distribution
can be generated {\em exactly} with the following algorithm:

\begin{enumerate}

\item We generate a random number in the interval $[0, n_\mr p\cdot M+n_\mr f$,
where $M$ is an integer used to increase the probability of choosing a p-chain,
which is introduced to at least partly account for the large spin entropy of
p-chains. If the integer generated is smaller than $n_\mr p\cdot M$, we choose
a p-chain, otherwise we manipulate an f-chain.

\item Suppose that the mth p-chain is chosen (with probability $\frac{M}{n_\mr
p\cdot M+n_\mr f}$), we then randomly single out a bead j in this chain (the
probability that the ith bead being chosen $1/N$), and subsequently we flip the
bead's identity.

If $\tau_\mr{mj}=0$, we attempt to change it to $\tau_\mr{mj}=1$. According to
the definition of the p-chains, in this case, the p-chain is still a p-chain
after the switch. We write the transition probability of such a switch as
\begin{equation}
    P^\mr{pp}(0\to 1)
        =\frac{M}{n_\mr p\cdot M+n_\mr f}\frac{1}{N} W^\mr{pp}_{01},
\end{equation}
where $W^\mr{pp}_{01}$ denotes the acceptance probability, which will
be specified below.

If $\tau_\mr{mj}=1$, we try to flip it to $\tau_\mr{mj}=0$. The chain type
(p-chain or f-chain) after the switch depends on the spin values of all other
beads. We have to distinguish between two different cases. 

In the first case, at least one other bead (say the lth bead) carries a spin
$\tau_\mr{ml}=1$. After the switch, the chain is then still a p-chain. The
transition probability is 
\begin{equation}
    P^\mr{pp}(1 \to 0)
        =\frac{M}{n_\mr p\cdot M+n_\mr f}\frac{1}{N} W^\mr{pp}_{10},
\end{equation}
where $W^\mr{pp}_{10}$ is again an acceptance probability.

In the second case, we have $\tau_\mr{ml}=0$ for all $l\neq j$, and the switch
will change this p-chain to an f-chain. The transition probability for such a
switch is written as
\begin{equation}
    P^\mr{pf}_\mr j [\{\mb R_\mr{m j}\}] 
        =\frac{M}{n_\mr p\cdot M+n_\mr f}\frac{1}{N} 
            W^\mr{pf}_\mr j 
%[\{\mb R_\mr{m j}\}] 
.
\end{equation}
Here, $P^\mr{pf}_\mr j$ and the acceptance probability $W^\mr{pf}_\mr j$ may
depend on the conformation $\{\mb R_\mr{m j}\}$ of the chain m.

\item If an f-chain has been picked in step 1 (with probability $\frac{n_\mr
f}{n_\mr p\cdot M+n_\mr f}$), we attempt to switch it to a p-chain. To do so we
first choose a bead j (with probability $1/N$) and a location $\mb r$ according
to a probability $H_\mr j(\mb r)$, and then generate a trial bead at the
position $\mb r$. In case the move is accepted, this bead becomes the j-th bead
of a p-chain and replaces the j-th bead of an f-chain. Hence the probability of
picking that f-bead is $H_\mr j(\mb r)/\rho_0 \phi_\mr{f,j}(\mb r) \Delta V$,
where $\Delta V$ is the volume of a cell, and $\rho_0 \phi_\mr{f,j}$ the
density of j-th monomers of f-chains. Then we construct a Gaussian chain with
jth bead fixed at $\mb r$.  The a priori probability $P_\mr{priori}$ to
generate a given set of coordinates $\{\mb R_\mr k \}$ is 
\begin{equation}
P_\mr{priori}[\{\mb R_\mb k\}] = \frac{1}{Z_0}\exp \Big[
	- \frac{N}{4} \sum_{j=2}^\mr N
	\big(\mb R_\mr{j}-\mb R_\mr{j-1}\big)^2\Big],
\end{equation}
where the normalization factor $\mc Z_0 = (\frac{N}{4 \pi})^{3(N-1)/2} $
corresponds to te partition function of an ideal noninteracting reference chain
of length $N$.  The transition probability from an f-chain to this p-chain is
thus expressed as
\begin{equation}
    P^\mr{fp}[\{\mb R_\mr k\}] 
	= \frac{n_\mr f}{n_\mr p\cdot M+n_\mr f}
          \frac{H_\mr j}{\rho_0 \phi_\mr{f,j} \Delta V}
	   P_\mr{priori} \: W^\mr{fp}_\mr j
%[\{\mb R_\mr k\}]
,
\end{equation}
with the acceptance probability $W^\mr{fp}_\mr j$, which again may
depend on the conformation of the new p-chain.

\item The trial move is accepted with probability $W^\mr{pp}_{01}$,
$W^\mr{pp}_{10}$, $W^\mr{pf}_\mr j$, or $W^\mr{fp}_\mr j$, respectively.
To ensure local mass conservation, this is done in a manner that the total
density $\hat{\phi}_\mr p (\mb r) + \phi_\mr f(\mb r)$ is preserved everywhere.
Hence, if a switch from the mth p-chain to the mth f-chain is accepted, we
remove its previous contribution to the density, say $\hat\phi_\mr m$ from
$\hat\phi_\mr p$, and add it to $\phi_\mr f$. This also entails the calculation
of the new $\omega_f$ which gives $\phi_\mr f+\phi_\mr m$.  Similarly, if the
switch from an f-chain to a p-chain is accepted, we remove the density
distribution of the newly generated p-chain from $\phi_\mr f$ and update
$\omega_\mr f$ accordingly.

\end{enumerate}

If only switch moves are made, the target spin distribution (\ref{eq:target})
can be obtained by choosing the acceptance probabilities $W^\mr{pp}_{01}$,
$W^\mr{pp}_{10}$, $W^\mr{pf}_\mr{mf}$, and $W^\mr{fp}_\mr j$ such that they
satisfy the detailed balance condition.  In the case of p-p switches, this
condition simply reads $ w_0 P^\mr{pp}_{01} = w_1 P^\mr{pp}_{10}$, and
it can be implemented with the simple Metropolis criterion
\begin{equation}
\label{eq:wpp}
	W^\mr{pp}_{01}=\min(1,e^{-\Delta\mu}), \;
	W^\mr{pp}_{10}=\min(1,e^{\Delta\mu}). 
\end{equation}

In the case of p-f or f-p switches, one must account for the entropy gain
associated with the replacement of an actual p-chain m with density
distribution $\hat{\phi}_\mr m$ by a f-chain density distribution $\phi_\mr m$.
The corresponding free energy difference is given by $ \Delta \mc F_0 = \Delta
\mc F_{0,\mr p} + \Delta \mc F_{0, \mr f}$ with $\Delta	\mc F_{0,\mr p} = -
\frac{N}{4} \sum_{j=2}^\mr N \big(\mb R_\mr{j}-\mb R_\mr{j-1}\big)^2$ and
$\Delta \mc F_{0,\mr f} = \mc F_{0,\mr f}[\phi_\mr f + \phi_\mr m] - \mc
F_{0,\mr f}[\phi_\mr f]$, where $\phi_\mr f$ denotes the density distribution
of f-chains without the chain $m$. Thus the detailed balance condition reads
\begin{equation}
w_1 P^\mr{pf}_\mr j =  w_0 P^\mr{fp}_\mr j \: e^{-\Delta \mc F_0} ,  
\end{equation}
which can be achieved with the Metropolis form
\begin{eqnarray}
\label{eq:wpf}
	W^\mr{pf}_\mr j &=&\min(1,e^{-\Delta\mu - \Delta \mc F_{0,\mr f}}
          \frac{H_\mr j(\mb r_\mr j)}
            {Z_0 \rho_0 \phi_\mr{f,j}(\mb r_\mr j) \Delta V}) 
\\
\nonumber
	W^\mr{fp}_\mr j &=&\min(1,e^{\Delta\mu + \Delta \mc F_{0,\mr f}}
	  \frac{Z_0 \rho_0 \phi_\mr j(\mb r_\mr j) \Delta V}
        {H_\mr j(\mb r_\mr j)}).
\end{eqnarray}
With this choice of acceptance probabilities, one would obtain the target spin
distribution (\ref{eq:target}) in a simulation that only includes spin
switches, i.e., chain identity switches. However, the expressions
(\ref{eq:wpf}) are quite complicated and involve time consuming calculations of
$\Delta \mc F_{0,\mr f}$ and $\phi_\mr{f,j}$. On the other hand, an exact
implementation of a target spin distribution is not truly necessary, since the
spins are auxiliary variables and have no physical meaning. Moreover the BD and
LD moves described in Secs \ref{subsec:bd} and \ref{subsec:ld} distort the spin
distribution anyway, since they do not satisfy detailed balance with respect to
(\ref{eq:target}). Therefore, we have chosen to simplify the expressions
(\ref{eq:wpf}) by making the following approximations: First, we replace
$\Delta \mc F_{0,\mr f} \approx -\ln(Z_0)$. Second, we neglect the fact that
$\rho_0 \phi_\mr{f,j}$ depends on j, i.e., we approximate $\rho_0 \phi_\mr{f,j}
\approx \frac{1}{N} \rho_0 \phi_\mr j$, and choose $H_\mr j(\mb r) \equiv H(\mb
r) = \rho_0 \phi_\mr f(\mb r)$ in step 3.  With these simplifications, the
acceptance probability (\ref{eq:wpf}) takes the simple form
\begin{equation}
\label{eq:wpf_simple}
	W^\mr{pf}= \min(1,e^{-\Delta\mu}), \;
	W^\mr{fp}=\min(1,e^{\Delta\mu}). 
\end{equation}
in analogy to Eq.\ (\ref{eq:wpp}). 

\section{Results and discussion}
\label{sec:results}

\subsection{Chain partitioning in a polymer solution}
\label{subsec:solution}

We first test the partitioning of p-chains and f-chains driven by the tuning
function in a homogeneous polymer solution with periodic boundary conditions.
We choose the total number of chains $n_\mr t=10^4$, and chain length $N=20$
for all chains.  Initially, we set $n_\mr p=n_\mr f=5000$, and the
configurations of p-chains are generated randomly as Gaussian coils, while the
field density is set to be homogeneous. The size of the box is $L_\mr x=L_\mr
y=2$, and $L_\mr z=16$ with discretization $n_\mr x=n_\mr y=8$, and $n_\mr
z=64$, i.e., we have uniform cubic cells with volume 0.125. The average number
of beads in each cell is about 50, which should be large enough to generate
smooth densities. The excluded volume parameter is set $v_\mr{ex}=1$.

To integrate the dynamic equations numerically, we choose a multiple time step
approach with time step $\Delta t_\mr{BD}=5\times 10^{-4}t_0/N$ for the BD
scheme, and $\Delta t_\mr{LD}=10^{-4}t_0$ for the LD equation. During each
evolution step, we thus update f-chains one time and p-chains 4 times, i.e. we
have $\Delta t\equiv\Delta t_\mr{LD}=4\Delta t_\mr{BD}$. 
% It should be noticed that $\Delta t_\mr{LD}$ can not be set to large,
% otherwise field density may become negative, which will terminate the
% simulation.  
In each simulation step, after updating the p-chains and f-chains, we attempt
to switch beads $S=500$ times, and about 10 chains are successfully switched.
The parameter used to enhance the probability of finding a p-chain is set to
$M=20$.  With these parameters, the density profiles for p-chains and f-chains
reach equilibrium within about half the relaxation time $t_0$ of a single
chain.

In principle, the tuning function can be chosen at will. In our test,
we assume that $\Delta\mu$ depends only on $z$ and has an almost
steplike profile,
\begin{equation}
    \Delta\mu(z)
        =\frac{\mu_\mr b+\mu_\mr m}{2}
        +\frac{\mu_\mr b-\mu_\mr m}{2}
        \tan\Big[\eta\cos\frac{2\pi}{L_\mr z}\Big(z-\frac{L_\mr z}{2}\Big)\Big]
\end{equation}
where $\mu_\mr b=0$, $\mu_\mr m=-1$, $\eta=10$ are parameters determining the
shape of $\Delta\mu$. 

%To clearly show the effect of tuning function, it is helpful to consider just
%a simple liquid updated with switches only. Define $\phi_1$ as the probability
%to find a p-particle, $\phi_2$ the probability to find an f-particle. For a
%homogeneous $\Delta\mu>0$, the detailed balance requires
%$\phi_1e^{-\Delta\mu}=\phi_2$, and considering also the normalization
%constraint $\phi_1+\phi_2=1$, we can obtain $\phi_1=(e^{-\Delta\mu}+1)^{-1}$
%and $\phi_2=1-\phi_1$. It is obviously that
%$\phi_1(\Delta\mu)=\phi_2(-\Delta\mu)$, and if we set $\Delta\mu=0$, we get
%$\phi_1=\phi_2=0.5$ as expected since the p-particles and f-particles are
%completely symmetric. The situation becomes complex when polymer chains are
%involved. For example, due to the asymmetric spin configuration space for
%p-chains and f-chains, the above symmetry does not hold, and analytical
%calculation becomes complex.
%
% FS: This paragraph somewhat disrupts the flow of the paper.
%     Let's add it when the referees ask for it ...

\begin{figure}[htb]
  \centering
    \includegraphics[angle=0, width=7.0cm]{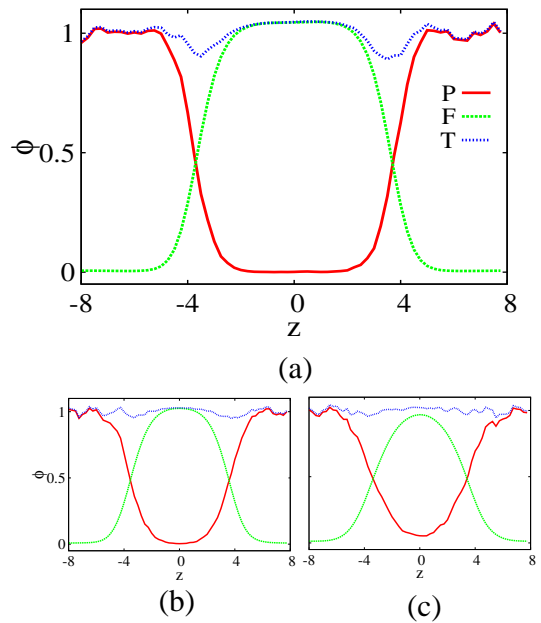}
  \caption{P-chain and f-chain density profiles showing the partitioning of
the system into particle-based regions and field-based region in simulations
of homopolymer solutions. Here P denotes
the density profile of p-chains, F the density for f-chains, and T
denotes their sum. For comparison, (b) and (c) shows the results for
different choices of the number $S$ of p-f switch attempts per time step:
(a) $S=500$ (default); (b) $S=50$; (c) $S=20$.
}
  \label{fig:solution} 
\end{figure}

Figure~\ref{fig:solution} shows the density profiles of p-chains, f-chains, and
their sum after equilibration of the system. As expected, p-chains aggregate
near the borders at $z=-8$ and $z=8$, while f-chains mainly appear in the
middle of the system. P-chains and f-chains are separated by interfaces with
width about 2.5$R_\mr g$. The interface region can become even narrower if the
tuning function is chosen sharper. However, due to the finite extension of
chains, it must have a minimum width around $R_\mr g$.  In principle, the total
density should be independent of $\Delta\mu$, i.e., the summation of p-chain
and f-chain density should be homogeneous. Indeed, in
Figure~(\ref{fig:solution}), the total density in the p-chain region and
f-chain region is almost constant apart from small statistical fluctuations.
However, in the interface region, a small dip appears. Such dips are often
observed in the interfacial regions of mixed resolution models
\cite{fritsch,poma,HPF1}. In our case, it can partly be related to a
discretization effect, as it becomes weaker if the number of grid points is
increased. If necessary, it could be removed by introducing additional
potentials in the interface region, following Ref.\ \cite{fritsch}.

The efficiency of partitioning of course depends on the frequency of chain
switch attempts. This is demonstrated in Figure~(\ref{fig:solution}) (b) and
(c), where the number of attempts has been reduced from the ``default'' value
$S=500$ (Figure ~\ref{fig:solution} a) to $S=50$ and $S=20$ (Figure
~\ref{fig:solution} b) and c). The interdiffusion of p- and f-chains competes
with the chain switching, and therefore, the interfacial region broadens. This
also reduces the density dip.

\subsection{Interface broadening in an A/B polymer blend}
\label{subsec:interdiffusion}

Next we test our hybrid model on a more complex problem, the interdiffusion of
miscible A/B polymers at an A/B interface.  For comparison, we also run pure
particle-based BD simulations (considered as the reference system) and pure LD
calculations for the same system.

We consider systems containing $n_\mr A=5000$ A polymers and $n_\mr B=5000$
B polymers with the same chain length $N=20$. 
%During the simulation $n_\mr A$ and $n_\mr B$ are kept fixed, i.e., we do not consider the switch between A polymers and B polymers. 
The melt is assumed to be compressible with $\kappa = 10$, and A/B monomers are
taken to be fully compatible, i.e., the incompatibility parameter $\chi$ (the
Flory Huggins parameter) is set to $\chi=0$.
Thus the nonbonded interactions are described by the Hamiltonian
\begin{equation}
\label{eq:HI_blend}
    \mc H_\mr I
    =\frac{n_\mr t\kappa}{2V}\int d\mb r\big(\phi_\mr A+\phi_\mr B-1\big)^2
\end{equation}
where $n_\mr t=n_\mr A+n_\mr B$ is the total number of polymers in the system.
This replaces (\ref{eq:HI}) in Sec \ref{sec:method}, and Eq. (\ref{eq:FI})
changes accordingly.

All simulations are implemented in three dimensional space with size $L_\mr
x=L_\mr y=2$, and $L_\mr z=24$ decomposed into $n_\mr x\cdot n_\mr y\cdot n_\mr
z=8\cdot 8\cdot 96$ uniform cells. To integrate the dynamical equations, we
choose $\Delta t_\mr{BD}=5\cdot 10^{-4}t_0/N$, and $\Delta t_\mr{LD}=10^{-4}$
as in Sec \ref{subsec:solution}.  Similarly, we also set $M=20$ and attempt 500
p-f switches are tried in each evolution step. Final statistical quantities are
obtained by averaging over the results from 64 separate independent runs.

\begin{figure}[hb]
  \centering
    \includegraphics[angle=0, width=7.0cm]{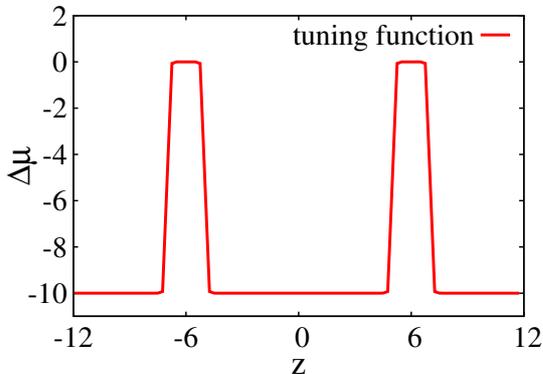}
  \caption{Profile of the tuning function, Eq.~(\ref{eq:mu_blend}). This 
 tuning function is adopted for both A polymer and B polymers.}
  \label{fig:delta_mu} 
\end{figure}

\begin{figure}[hb]
  \centering
    \includegraphics[angle=0, width=7.0cm]{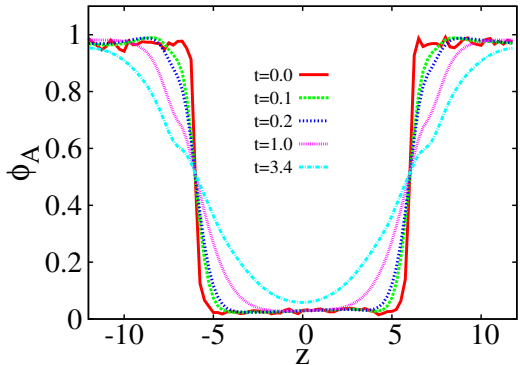}
  \caption{Density profiles for A polymers at different times from hybrid simulations. The times are measured in units of $t_0$.}
  \label{fig:evolution} 
\end{figure}

Initially, we set the number of p-chains and f-chains for A and B polymers to
equal values, i.e., $n_\mr{Ap}=n_\mr{Af}=n_\mr{Bp}=n_\mr{Bf}=2500$. The initial
configuration for p-chains are generated from additional MC simulations leading
to homogeneous density profiles along the $x$ and $y$ directions and sharp
interfaces along $z$ direction. Therefore, in the following, we only consider
the density profiles along the $z$ direction. For simplicity, the initial
density profiles of f-chains are set such that
$\phi_\mr{Af}(t=0)=\hat\phi_\mr{Ap}(t=0)$, and
$\phi_\mr{Bf}(t=0)=\hat\phi_\mr{Bp}(t=0)$. At $t=0$, we thus have a strongly
phase separated system with an A rich region near the boundaries (located at
$z=-12$, and $z=12$), and a B rich region in the middle (See
Fig.~(\ref{fig:evolution}) for the initial density profile of A polymers
$\phi_\mr A(t=0)$. The profiles of $\phi_\mr B(t=0)$ are not shown).

\begin{figure}[ht]
  \centering
  \subfigure[]{
    \label{fig:3a} 
    \includegraphics[angle=0, width=7.0cm]{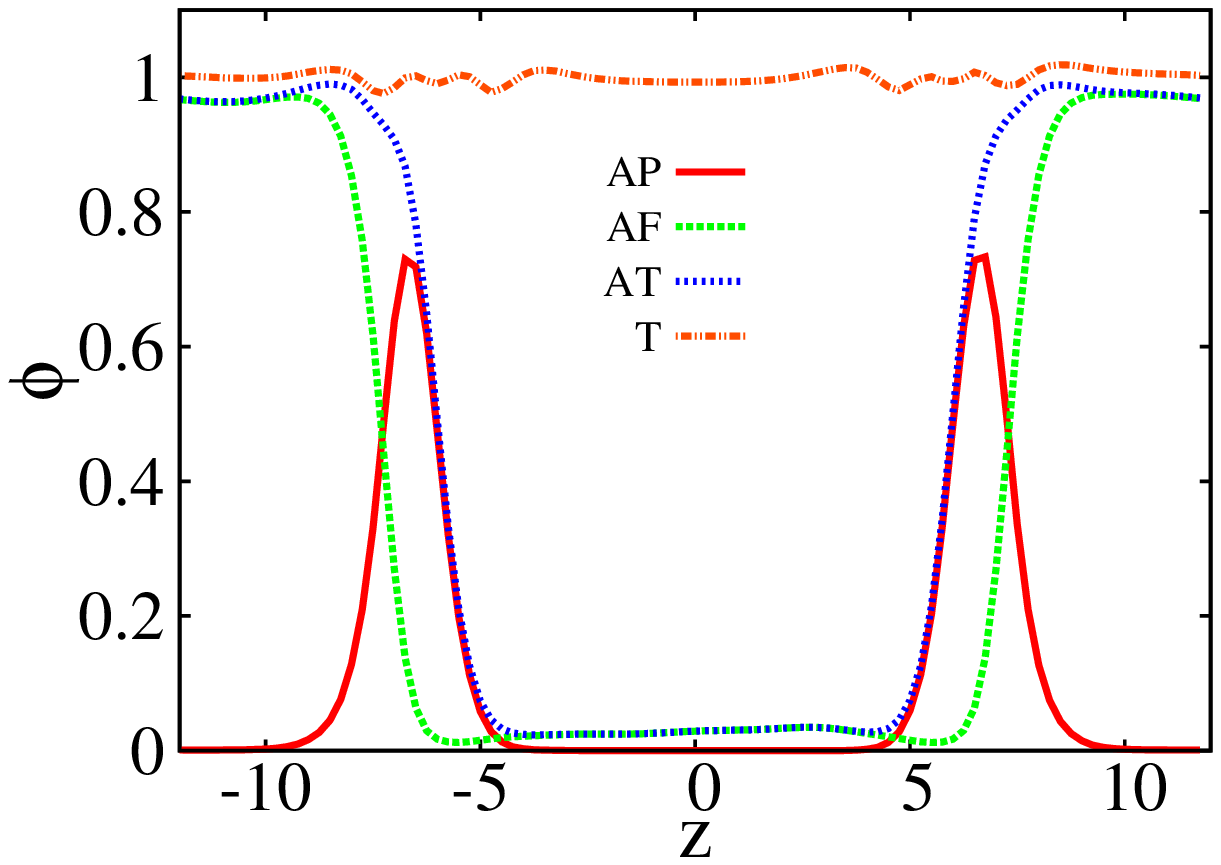}}
  \subfigure[]{
    \label{fig:3b} 
    \includegraphics[angle=0, width=7.0cm]{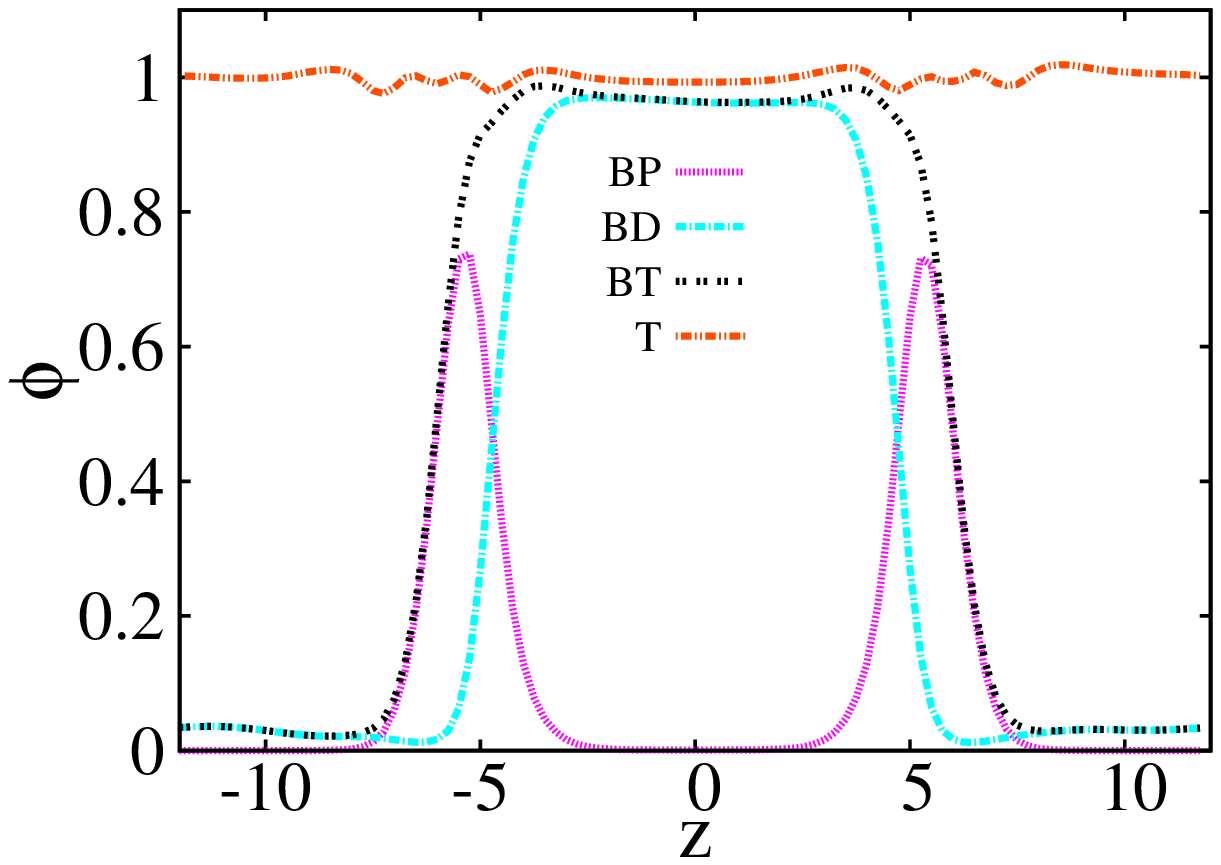}}
  \caption{Density profiles for p-chains (AP), f-chains (AF), and the total of A polymers (AT) (a); density profiles for p-chains (BP), f-chains (BF), and the total of B polymers (BT) (b). Here T denotes the profile of total density of A and B polymers. These density profiles are extracted from the system at $t=0.1t_0$
}
  \label{fig:blend} 
\end{figure}

We are interested in the A/B interface regions and want to keep the chain
configuration properties there. Thus we need to resolve polymers as particles
in the interface regions, while we can represent them by fields elsewhere. This
can be done by choosing a tuning function for both A and B polymers such that
it has large values in the interface regions and small values elsewhere.
Specifically, we use
\begin{equation}\label{eq:mu_blend}
\Delta\mu=\frac{\mu_\mr a+\mu_\mr b}{2}+\frac{\mu_\mr a-\mu_\mr b}{2}\tan\bigg[\eta\Big(\Big||z|-\frac{L_\mr z}{4}\Big|-\Delta\Big)\bigg]
\end{equation}
where $\mu_\mr a=-10$, $\mu_\mr b=0$, $\eta=10$, $\Delta=1$ are parameters
controlling the shape of $\Delta\mu$. The tuning function $\Delta\mu$ is 
plotted in Figure~\ref{fig:delta_mu}.

Figure~\ref{fig:blend} shows the density profiles at $t=0.1t_0$ of p-chains and
f-chains for A and B polymers representing the successful partitioning of
chains according to the tuning function. These one dimensional densities are
obtained by averaging over the corresponding three dimensional densities in the
$x$ and $y$ directions. In the early stage, $t\sim 0.1t_0$, the shape of the
initial densities does not change very much, and most of the A polymers are
still located near boundaries, while the B polymers are confined to a slab in
the middle of the system. As expected, p-chains of both A and B polymers are
mainly found in the interface regions (around $L_\mr z=-6$ and $L_\mr z=6$)
with width about 2.5$R_\mr g$, and f-chains are distributed elsewhere. We find
that that at $t\sim 0.1t_0$, the number of p-chains is roughly $n_\mr{Ap}\sim
n_\mr{Bp}\sim 1000$, while the number of f-chains is $n_\mr{Ap}\sim
n_\mr{Bp}\sim 4000$, i.e., about 20\% of A polymers (and B polymers) are
resolved by BD particles. Small dips can also be seen in the total density
profiles similar to that in polymer solutions.  With the presently chosen
parameters, f-chains are not fully replaced by p-chains in the A/B interface
regions, i.e., the maximum value of $\phi_\mr{Ap}$ and $\phi_\mr{Bp}$ does not
reach the bulk value.  One could easily adjust the parameters such that a wide
slab in the interfacial region is filled by p-chains only. However, here we are
interested in the interplay of particles and fields and its influence on the
interdiffusion, which is why we keep the p-domains very narrow and f-chains can
intrude.

Next we investigate the interfacial broadening process at early stages. This is
done by monitoring the evolution of polymer densities. Since A polymer and B
polymers are symmetric, we just focus on the properties of A polymers.
Figure~\ref{fig:evolution} shows the evolution of the A polymer density
$\phi_\mr A(t)=\hat\phi_\mr{Ap}(t)+\phi_\mr{Af}(t)$. Starting from the initial
sharp density profile, $\phi_\mr A(t)$ gradually becomes diffuse. At late times
(not hown), $\phi_\mr{A}(t)$ approaches a uniform distribution with magnitude
about 0.5.

\begin{figure}[hb]
  \centering
    \includegraphics[angle=0, width=7.0cm]{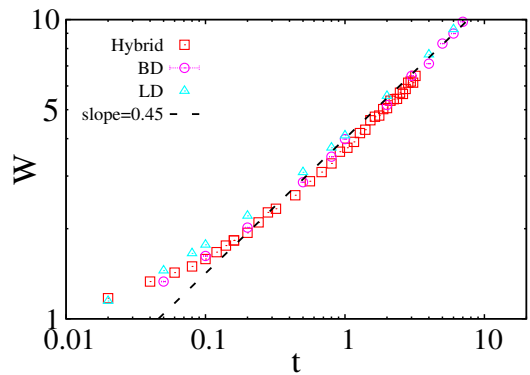}
  \caption{Evolution of A/B interfacial width $W$ as a function of time (in units $t_0$) according to different models as indicated. The dashed line is plotted to guide eyes and it indicates a power law $W(t)\sim t^a$ with $a\simeq 0.45$. The size of error bar for the data obtained from hybrid simulations is similar to that from pure BD simulations (not shown).}
  \label{fig:width} 
\end{figure}

To characterize the interfacial broadening more quantitatively, we introduce a
quantity $W$ measuring the interfacial width, and monitor $W$ as a function of
time. The ``interfacial width'' $W$ is simply defined as the inverse of the
maximum slope of the density profile. We do not renormalize it to the ``bulk
densities", as the latter are changing with time and are thus not well defined.
Figure~\ref{fig:width} plots the evolution of width $W(t)$ obtained with our
hybrid method, and compares with with data from pure particle-based BD
simulations and from pure field-based LD simulations.  The initial densities in
the pure LD calculations is set the same as that in the pure BD simulations.
Except for early times ($t<0.1t_0$), where deviations are observed due to the
finite interfacial width effect, the curves for the hybrid model, the pure BD
simulations and pure LD theory are rather well described by a scaling relation
\cite{broaden1,broaden2,broaden3} $W\sim t^a$ with exponent $a\sim0.45$.

The simulation data obtained with the hybrid model are in excellent agreement
with those obtained from the fully fine-grained BD model. In contrast, the data
from the LD simulations show deviations at early times. These presumably
reflect the importance of density fluctuations in the interface region, which
are neglected in the coarse-grained LD approach. This problem can be avoided by
resorting to the hybrid description.

\section{Summary and remarks}
\label{sec:summary}

In this work, we have developed a hybrid scheme that dynamically couples a
particle-based description of polymers to a field-based description in an
adaptive resolution sense, where polymers can switch between different
resolutions on the fly. The resolution switch is controlled by a predefined
tuning function through a Monte Carlo criterion.  In order to capture the
dynamic behavior, we use Brownian dynamics at the particle level, and a local
density functional theory at the field level. The hybrid scheme was explained
in detail at the example of a polymer solution, and then tested in simulations
of a more complex problem, the interface broadening in a polymer A/B blend.

In the applications presented here, the tuning function was kept constant
throughout the simulation. The geometric partitioning into particle and
field domains was thus pre-defined prior to the simulation. However, the
partitioning has no physical meaning, and the tuning function can be chosen
at will, which means that it can also be adjusted to the configuration during
the simulation. For example, in a large-scale simulation of A/B interfaces 
in immiscible polymer blends, capillary undulations of the interface must
be taken into account. The fine-grained p-domain should thus be able to
follow the local position of the interface. This could be done, e.g., by 
coupling the tuning function to the local concentration difference of
A and B monomers, or to concentration gradients.

We should note apart from LD, a number of other dynamic density functional
theories (DDFTs) have been proposed for dynamic studies of polymer systems
\cite{LD2,fluctuation_dynamics}. Unfortunately, in contrast to simple liquids
\cite{simple_DFT1,simple_DFT2,simple_DFT3,simple_DFT4,simple_DFT5}, it is not
possible to derive d DDFT for polymers which is based on densities only and
fully equivalent to the BD description \cite{BD_to_DDFT1,BD_to_DDFT2}.
Therefore, DDFTs for polymers always rely on approximations and represent
different aspects of the polymer motion. The local DDFT used here (the LD
method) focusses on individual monomers and disregards the effect of chain
connectivity. Other DDFTs such that Debye dynamics and external potential
dynamics \cite{LD2} account for the chain connectivity and focus on the
collective motion of monomers in a chain.  In a recent work \cite{BD_and_DDFT},
we have compared several DDFT approaches with BD simulations, and shown that
the LD scheme reproduces the interfacial evolution in A/B homopolymer or the
kinetics of microscopic phase separation in A:B diblock copolymer melts fairly
accurately. However, in other situations, the kinetic pathways of structure
formation in DDFT simulations may depend on the choice of the DDFT
\cite{LB_DSCF, LB_EPD}. Therefore, to complete the hybrid modelling scheme, we
also tested models where the f-chains are propagated according to DDFT
equations that account for chain connectivity, e.g., Debye dynamics.
Unfortunately, we found that such schemes tend to suffer from a flow imbalance
in the f-p interfacial region. As a result, the overall density is no longer
constant, an accumulation of chains is observed in the p- or f-domain, and
corrections have to be applied.  A detailed analysis will be published
elsewhere.

Another important extension in future work will be to add inertia and account
for momentum conservation. At the coarse-grained particle level, this is easily
done by switching from Brownian Dynamics to Molecular Dynamics, possibly in
combination with a momentum-conserving thermostat that restores friction
effects \cite{agur}. At the field level, one option is to use a recently
developed scheme that couples DDFT with Lattice Boltzmann simulations
\cite{LB_DSCF, LB_EPD}.

Hybrid particle-field schemes can be useful in a variety of contexts: Firstly,
they can be used to implement proper boundary conditions in field-based
simulations \cite{HPF2}. This is particularly important in situations where
complex surfaces or interfaces critically influence the relevant properties of
a material. Examples are liquid crystal cells where surfaces define the
orientation of the director, or composite materials with polymer-coated filler
particles. Secondly, they can be used for particle- based simulations of
polymeric systems with open boundary conditions. The main purpose of the field
domain is then to provide a realistic environment for the system of interest
and act as polymer reservoir. We believe that our HPF approach and suitable
refinements and extensions will provide a convenient and practical tool for
such applications. 

\bigskip
\begin{center}
\textbf{ACKNOWLEDGMENTS}
\end{center}

We wish to thank H. Behringer and T. Raasch for useful discussions. 
Financial support from the German Science Foundation (DFG) within project C1 in
SFB TRR 146 is gratefully acknowledged. Simulations have been carried out on the
supercomputer Mogon at Johannes Gutenberg University Mainz (hpc.uni-mainz.de).

\end{document}